\documentclass[prb,aps,epsf,twocolumn,showpacs,10pt]{revtex4-1}
\usepackage{graphicx,amsfonts,times,bm,amsmath,verbatim,color,array}

\newcommand{\ket}[1]{| {#1} \rangle}
\newcommand{\expect}[1]{\langle {#1} \rangle}

\begin{document}

\title{Equivalence of topological mirror and chiral superconductivity in one dimension}
\author{Eugene Dumitrescu$^1$}
\author{Girish Sharma$^1$}
\author{Jay D. Sau$^2$}
\author{Sumanta Tewari$^1$}

\affiliation{ $^1$Department of Physics and Astronomy, Clemson University, Clemson, SC 29634\\
$^2$ Condensed Matter Theory Center, Department of Physics, University of Maryland, College Park, MD 20742}

\begin{abstract}
Recently it has been proposed that a unitary topological mirror symmetry can stabilize multiple zero energy Majorana fermion modes in one dimensional (1D) time reversal (TR) invariant topological superconductors.
Here we establish an exact equivalence between 1D ``topological mirror superconductivity'' and chiral topological superconductivity in BDI class which can also stabilize multiple Majorana-Kramers pairs in 1D TR-invariant topological superconductors. The equivalence proves that topological mirror superconductivity can be understood as chiral superconductivity in the BDI symmetry class co-existing with time-reversal
symmetry. Furthermore, we show that the mirror Berry phase coincides with the chiral winding invariant of the  BDI symmetry class, which is independent of the presence of the time-reversal symmetry. Thus,
the time-reversal invariant topological mirror superconducting state may be viewed as a special case of the BDI symmetry class in the well-known Altland-Zirnbauer periodic table of free fermionic phases.
We illustrate the results with the examples of 1D spin-orbit coupled quantum wires in the presence of nodeless $s_{\pm}$ superconductivity and the recently discussed experimental system of ferromagnetic atom (Fe) chains embedded on a lead (Pb) superconductor.
\end{abstract}

\maketitle

{\em Introduction :}
Mirror symmetry, when coupled with time-reversal (TR) and particle-hole (PH) symmetries, has recently been proposed\cite{Zhang_13} to stabilize a one-dimensional topological superconducting phase with an integer number of spatially overlapping zero energy Majorana fermion modes.
The number of Majorana edge modes in a so-called ``topological mirror superconductor" is indexed by a mirror Berry phase $\gamma_m \in \mathbb{Z}$ which was defined in Ref.~\citenum{Zhang_13}.
Mirror symmetry and a mirror Berry phase topological invariant have been used to theoretically explain the existence of multiple Majorana modes in heterostructure based topological superconductors involving a proximity induced $s_\pm$-wave pairing potential on spin-orbit coupled semiconductor wires\cite{Zhang_13}. A mirror symmetry has also been recently invoked to discuss the topological properties in the context of ferromagnetic atomic chains embedded on the surface of a Pb superconductor \cite{Yazdani_14,Li_14}.

Topological classification based on a unitary mirror symmetry aims to extend the $\mathbb{Z}_2$-invariant of the DIII symmetry class  to allow the existence of multiple Majorana Kramers pairs (MKPs)\cite{Zhang_13}.
In the Altland-Zirnbauer framework of classification of free fermionic phases \cite{AZ, Kitaev_2009, Ryu_2010}, Majorana fermions have been predicted to occur as edge states in low dimensional systems belonging to the topological classes D \cite{Tewari-strontium, Fu-Kane, Zhang-Tewari,Sato-Fujimoto, Sau, Long-PRB, Roman, Oreg}, DIII \cite{Nakosai_13, Sato_11, Law_12, Deng_12, Keselman_13, Liu_14, Deng_13, Zhang_13_TR, Flensberg_14, Arbel_14, Dumitrescu_TR}, and BDI\cite{TS_BDI, minigap, Chakravarty,Diez,Wong_13,He,Beenakker_14,Dumitrescu}.
This raises the fundamental question of whether or not mirror symmetry in one dimension refines the topological classification in the Altland-Zirnbauer framework, and if so, is mirror symmetry a critical
ingredient for the protection of MKPs  in time-reversal invariant DIII systems?
In one dimension it is well known that the chiral BDI symmetry class also supports an arbitrary number of Majorana end modes due to its $\mathbb{Z}$ invariant \cite{TS_BDI,minigap,Beenakker_14} and, with concurrent TR symmetry, stabilizes multiple Majorana-Kramers pairs \cite{Dumitrescu_TR} as edge modes.
It is therefore natural to ask how these two symmetries (mirror and chiral), which both produce multiple MKPs in the presence of concurrent TR,  are related and investigate the conditions under which they may differ, if at all.

In this work we formulate a general procedure illustrating how the presence of mirror symmetry promotes a one-dimensional class DIII topological superconductor to the chiral topological class BDI that co-exists
with a time-reversal symmetry. We then establish an exact equivalence between ``mirror Berry phase'' \cite{Zhang_13} used to justify the
presence of multiple MKPs in DIII systems and the integer winding number invariant of the BDI class \cite{Dumitrescu_TR}. However, the winding invariant for the BDI class
continues to be well-defined and unchanged even when the time-reversal symmetry is broken.
 The equivalence proves that ``topological mirror superconductivity'' \cite{Zhang_13} can be viewed as belonging to the BDI symmetry class in the well-known periodic table of Altland-Zirnbaur classification of free fermionic phases \cite{AZ, Kitaev_2009, Ryu_2010} and that the co-existing time-reversal symmetry does not qualitatively modify the topological phase. Therefore, in one dimension the chiral BDI symmetry is in fact more general than mirror symmetry in the sense that perturbations which break both the symplectic TR symmetry and topological mirror symmetry may keep the chiral symmetry invariant, allowing multiple Majorana fermion modes \textit{even in the absence of mirror symmetry}.
In contrast, chiral symmetry breaking terms also necessarily break either mirror or time reversal symmetry (under either of these circumstances the mirror Berry phase cannot be defined \cite{Zhang_13}) and therefore topological mirror symmetry cannot be thought of as an independent symmetry in one-dimension.

{\em Mirror versus chiral symmetric systems:}
In this paper we will consider a superconducting system with particle-hole symmetry $\Xi$  and time-reversal symmetry $\Theta$. Superconducting systems with mirror symmetry for which the ``mirror Berry phase'' is defined \cite{Zhang_13} have an additional unitary symmetry
operator $\cal M$ such that  $[{\cal M},\Theta]=[{\cal M},\Xi]=[{\cal M},H]=0$ (where $H$ is the BdG Hamiltonian for which explicit examples are discussed in later sections) and ${\cal{M}}^2=-1$. On the other hand, systems in
the chiral BDI symmetry class have a pseudo time reversal operator ${\cal O}$ (with ${\cal O}^2=1$) which in combination with the particle-hole symmetry $\Xi$ defines a chirality operator
${\cal C}=\Xi\cdot{\cal O}$ with the properties  ${\cal C}^2=1 ,\{{\cal C},H\}=0, [\Theta,{\cal C}]=0$. Note that we have assumed here that the two time-reversal operators ${\cal O}$ and $\Theta$
commute. From these definitions it is clear that when these BDI symmetry operators are present one can define a ``mirror'' operator ${\cal M}={\cal O}\cdot\Theta$, which satisfies all the properties
mentioned for the Mirror operator. Alternatively for systems with time-reversal and mirror symmetry one can define the pseudo-time reversal ${\cal O}=\Theta \cdot {\cal M}$, which is required to characterize systems in the
BDI symmetry class. Therefore, superconducting systems with TR and BDI symmetries have the same operator content as systems with TR and mirror symmetries.

{\em Equivalence of mirror Berry phase and BDI winding number:}
The topological invariant associated with the BDI symmetry i.e. the winding number for ${\cal C}$ coincides with the mirror Berry phase defined in Ref.~[\citenum{Zhang_13}] are also related.
To see this, we follow  Ref.~[\citenum{Zhang_13}] to define the mirror Berry phase and define a family of Hamiltonians parametrized by $\theta$ as
\begin{align}
& {H}(k,\theta)=H(k)\cos{\theta}+ {\cal C}\sin{\theta}.
\end{align}
Note here that we have introduced a slight modification to the definition used in Ref.~[\citenum{Zhang_13}], where for the chirality operator we have used ${\cal C}=\Pi \cdot {\cal M}$ instead of $\Pi=i\Theta\cdot\Xi$. But
since ${\cal M}$ commutes with $H$ this is essentially equivalent apart from the fact that with this transformation, the mirror Berry phase in Ref.~[\citenum{Zhang_13}]  becomes
the total Chern number of ${H}(k,\theta)$ .
 The Chern number is given by integrating the curvature of the Berry connection and using Stokes theorem is related to
the integral of the Berry connection written as
\begin{align}
&\bm a(k,\theta)=\sum_n f_n \expect{n,k,\theta|\bm\nabla_{\theta,k}|n,k,\theta}
\end{align}
at $\theta=0$,
where $\ket{n,k,\theta}$ are wave-functions parametrized either on the top Hemisphere $\pi/2>\theta>0$ or
the bottom Hemisphere  $-\pi/2<\theta<0$.

The Hamiltonian $H(k)$ is off-diagonal with an off-diagonal matrix $Q(k)$ as
\begin{align}
&H(k)=\left(\begin{array}{cc}0&Q(k)\\Q(k)^\dagger &0\end{array}\right)
\end{align}
in the basis where ${\cal C}$ is diagonal and has a  winding number ${\cal W}$ defined in terms of the phase $Arg(det(Q(k)))$ \cite{TS_BDI}. The matrix
$Q(k)$ can also be decomposed (singular value decomposition) as
\begin{align}
&Q(k)=U_k^\dagger \Sigma_k V_k,
\end{align}
where $U_k,V_k$ are unitary and $\Sigma_k$ is a positive diagonal matrix. The Hamiltonian $H(k)$
can be transformed using a unitary transformation
\begin{align}
&Z(k)=\left(\begin{array}{cc}U_k&0\\0&V_k\end{array}\right)
\end{align}
to define another chiral Hamiltonian
\begin{align}
&\bar{H}(k)=Z(k)H(k)Z(k)^\dagger=\left(\begin{array}{cc}0&\Sigma(k)\\\Sigma(k) &0\end{array}\right).
\end{align}
Since this Hamiltonian $\bar{H}(k)$ has wave-functions $\ket{n,k}_0$ that are easy to write down in terms of the diagonal matrix $\Sigma(k)$
one can easily check that this Hamiltonian has  vanishing winding number $0$ and also zero Chern number $0$.
The Berry connection of the two chiral Hamiltonians are related by
\begin{align}
&\bm a(k,\theta=0)=\sum_n f_n \expect{n,k|\bm\nabla_{k}|n,k}\\
&=\sum_n f_n \expect{n,k|Z(k)^\dagger\bm\nabla_{k}[Z(k)|n,k}_0]\\
&=\bm a_0(k,\theta=0)+\sum_n f_n \expect{n,k|[Z(k)^\dagger\bm\nabla_{k}Z(k)]|n,k}_0\\
&=\partial_k Arg[Det(U(k)V^\dagger(k))]=\partial_k Arg[Det[Q(k)]].
\end{align}
Therefore the integral of the Berry connection, which is related to the mirror Berry phase, is the same
as the winding number invariant of the BDI class.

{\em Example 1: Spin-orbit coupled nanowire proximity coupled to $s_{\pm}$ superconductor:}
As a concrete example, we illustrate the equivalence of topological mirror and chiral superconductivity for a system consisting of a spin-orbit coupled semiconductor nanowire with proximity induced $s_{\pm}$-wave superconductivity \cite{Zhang_13}.
In principle this may be experimentally achieved by depositing an InSb nanowire onto an Iron based superconductor with a sign changing extended $s$-wave order parameter.
The effective Bogoliubov-de Gennes (BdG) Hamiltonian for the nanowire with proximity induced superconductivity is $H=\sum_{k} \Psi_k^{\dagger} H (k) \Psi_k$ where,
\begin{eqnarray}
\label{eq:spm}
H (k) & = &  (-2t \cos(k) -\mu) {\sigma_0}\tau_z +\alpha_R \sin(k)  {\sigma_y} \tau_z\\
      & + & \Delta_s \cos(k) {\sigma_0} \tau_x \nonumber.
\end{eqnarray}
Here $k \equiv k_x$ is the 1D crystal momentum and $\Psi_{k}=(c_{k\uparrow},c_{k\downarrow},c_{-k\downarrow}^{\dagger},-c_{-k\uparrow}^{\dagger})^{T}$ is a four component Nambu spinor acting in the ${\tau}$ (particle-hole) and ${\sigma}$ (spin) spaces.
Additionally, $t$ is the nearest neighbor hopping, $\mu$ is the chemical potential, $\alpha_R$ is the strength of Rashba spin orbit coupling (which we haven chosen to be along $\hat{y}$ without loss of generality) and $\Delta_s$ is the proximity induced pair potential.

As with any superconducting mean field BdG Hamiltonian Eq.~(\ref{eq:spm}) is invariant under a particle-hole transformation denoted by the operator $\Xi$.
The PH constraint for Bloch Hamiltonians is $\Xi H (k) \Xi^{-1}=-H (-k)$ and the anti-unitary PH operator in our basis is given by $\Xi=\sigma_y \tau_y {\cal{K}}$.
In addition to belonging in the time reversal symmetry class DIII with a $\mathbb{Z}_2$ invariant (by virtue of the time-reversal symmetry $\Theta H (k) \Theta^{-1}= H (-k)$ where $\Theta=i \sigma_y \tau_0 {\cal{K}}$ with ${\cal K}$ the complex conjugation operator), the Hamiltonian in Eq.~(\ref{eq:spm}) is also invariant under a mirror symmetry\cite{Zhang_13} $\mathcal{M}=-i\sigma_y\tau_0$ since $\mathcal{M} H(k) \mathcal{M}^{-1} = H(k)$ (i.e. $\left[\mathcal{M} ,H(k)\right]=0$).
Note that the momentum has not changed sign under the mirror operation since the mirror operator $\mathcal{M}$ is unitary and the reflection associated with the mirror symmetry is taken about a 1D mirror line\cite{Zhang_13}.

Since the Hamiltonian in Eq.~(\ref{eq:spm}) is invariant under $\Theta$ and $\mathcal{M}$ as defined above, it is also clearly invariant under their composition which we define as $\mathcal{O} = \Theta \cdot \mathcal{M}$.
Explicitly, $\mathcal{O}= \sigma_0 \tau_0 {\cal{K}}$ which squares to $+1$ and the Hamiltonian in Eq.~\ref{eq:spm} transforms under this operator according to $\mathcal{O} H (k) \mathcal{O}^{-1}= H (-k)$, which is the pseudo time-reversal symmetry introduced above.
The presence of a TR-operator with $\mathcal{O}^2=1$, along with PH symmetry, means that Eq.~(\ref{eq:spm}) also satisfies the requirements to be in the topological class BDI indexed by an integer winding invariant $\mathcal{W} \in \mathbb{Z}$.
As we showed in the last section, the mirror Berry phase calculated in the presence of ${\cal M}$ and $\Theta$ is identical to the BDI winding number $\mathcal{W}$ that can be defined with the help of the pseudo TR operator ${\cal O}=\Theta\cdot{\cal M}$ (with ${\cal O}^2=1$) and the PH operator $\Xi$.
Below we first review the action of $\mathcal{M}$ and the pseudo TR operator $\mathcal{O}$ on the Pauli matrices $\sigma_i$, and then consider perturbations which remove each symmetry individually.

\begin{figure}[t!]
\centering
\includegraphics[width=6cm]{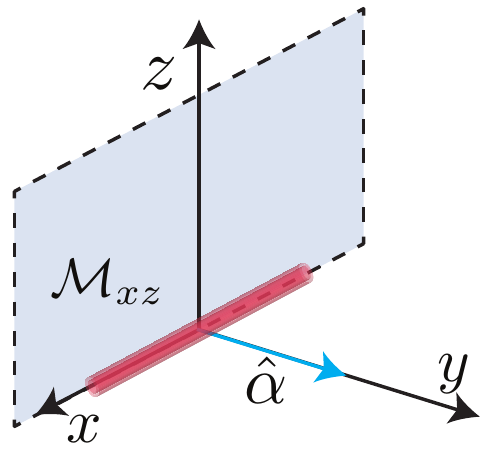}
\caption{Mirror symmetry for spin-orbit coupled wires (Example 1).
In this case the 1D nanowire (red) lies along the $x$-axis.
The blue arrow denotes the direction of the spin-orbit field in spin space which defines the mirror plane (light blue).
When a Zeeman field lies in the mirror plane the mirror symmetry is broken but chiral symmetry is preserved.
If a Zeeman field is normal to the mirror plane the mirror symmetry is preserved while time reversal and chiral symmetries are removed.
However, in this case, there is no mirror topological invariant since the mirror Berry phase is only defined in the presence of time-reversal symmetry (class DIII). Figure for mirror symmetry for ferromagnetic atom chain embedded on Pb superconductor (Example 2) should be analogous.}
\label{fig:mirror}
\end{figure}

It is straightforward to observe the following relations involving the action of TR (conventional as well as pseudo) and mirror symmetries on the Pauli matrices:
\begin{subequations}
\label{equations}
\begin{align}
\label{eq:paulia}
\Theta \sigma_i \Theta^{-1} & =   - \sigma_i  \\
\label{eq:paulib}
\mathcal{M} \sigma_i \mathcal{M}^{-1} & =  + \eta_{i} \sigma_i  \\
\label{eq:paulic}
\mathcal{O} \sigma_i \mathcal{O}^{-1} & =  - \eta_{i} \sigma_i
\end{align}
\end{subequations}
where $\eta_i = -1$ for $i=x,z$ and $\eta_i = +1$ for $i=y$ (note that in obtaining Eq.~\ref{eq:paulic} we have used the fact that $\mathcal{O} = \Theta \cdot \mathcal{M}$).
Thus, under the mirror operation ($\mathcal{M}$) spin components in the $x-z$ plane acquire a phase of $-1$ while the spin component normal to the mirror plane (see Fig.~\ref{fig:mirror}) is unchanged.
Note that the SU(2) angular momentum algebra of the $\sigma$ operators ($\left[\sigma_i, \sigma_j \right] = 2 i \epsilon_{ijk} \sigma _k$) is preserved under the action of $\mathcal{O}$ since two (one) operators are even (odd) under $\mathcal{O}$ \cite{Winkler}.

Consider case (i) -- A perturbation which potentially can break the mirror symmetry of Eq.~(\ref{eq:spm}) is a Zeeman term $H_Z = \mathbf{V}\cdot\mathbf{\sigma}$ when $|V_{xz}|\neq 0$ where $|V_{xz}|=\sqrt{V_x^2+V_z^2}$, and $\mathbf{V}=(V_x,V_y,V_z)$.
Such a perturbation also breaks the conventional TR symmetry $\Theta$ but keeps the pseudo TR symmetry (${\cal M}\cdot\Theta$) and hence the chiral symmetry of the total Hamiltonian intact so long as $V_y=0$. Thus, in this case, even though the mirror Berry phase can no longer be defined, the BDI chiral invariant ${\cal W}$ remains well defined and the system can host an integer number of protected Majorana fermion modes at the edges.
Now consider case (ii) -- when $\mathbf{V}=(0,V_y,0)$ the Hamiltonian is mirror symmetric (${\cal M}$ is unbroken) but chiral and TR symmetries both break down (i.e. the Hamiltonian is not invariant under $\Theta$ and, consequently, also not under ${\cal O}={\cal M}\cdot\Theta$).
In this case, even though the mirror symmetry persists, the mirror Berry phase procedure is no longer applicable due to the lack of time reversal symmetry (note that the mirror Berry phase is only defined in the presence of $\Theta$ \cite{Zhang_13}).
Thus we find that the mirror symmetry helps us find a suitable pseudo TR operator $\mathcal{O}$ which in turn helps us define the chiral operator $\mathcal{C}=\mathcal{O}\cdot\Xi$. This procedure is valid as long as the TR operator $\Theta$ is a symmetry or as in case (i) wherein both $\Theta$ \textit{and} ${\cal M}$ are broken by the same Zeeman field. Such a procedure of defining a chiral symmetry does not work in case (ii) and also in case (iii) where the mirror symmetry is broken (say, by an additional next-nearest-neighbor spin-orbit coupling $H_k^{SO}=\alpha^{\prime}\sin(2k)(\mathbf{c}\cdot\mathbf{\sigma})\tau_z$ with
$\mathbf{c}\perp \hat{y}$ \cite{Dumitrescu_TR}) but $\Theta$ is unbroken. Note that in both case (ii) and (iii) neither mirror Chern nor the chiral winding number invariant can be defined. However, as we have established in this work, when both invariants are defined (i.e., in the presence of $\Theta$), they are identical, and the topological mirror superconductivity and chiral BDI superconductivity are equivalent descriptions of the same physical system.

%
\begin{table}[t!]
\newcommand\T{\rule{0pt}{3.ex}}
\newcommand\B{\rule[-1.7ex]{0pt}{0pt}}
\centering
\begin{ruledtabular}
\begin{tabular}{cccc}
      Operator  & Symmetry & Ex. 1 & Ex. 2 \\[3pt]
      \hline
      $\Xi$ & PH & $\sigma_y \tau_y {\cal{K}}$   & $\sigma_y \tau_y {\cal{K}}$  \T\\[3pt]
      \hline
      $\mathcal{M}$ & Mirror & $i \sigma_y \tau_0 $   & $i \hat{d} \cdot \bm{\sigma} \tau_0$  \T\\[3pt]
      \hline
      $\Theta$ & DIII TR & $i \sigma_y \tau_0 {\cal{K}}$   & $i \sigma_y \tau_0 {\cal{K}}$  \T\\[3pt]
      \hline
      $\mathcal{O}$ & BDI TR & $ {\cal{K}}$   & $(\hat{d}\cdot\hat{y}+i(\hat{d}\times\hat{y})\cdot\mathbf{\sigma})\mathcal{K}$  \T\\[3pt]
      \hline
      $\Pi$ & DIII Chiral  & $ \sigma_0 \tau_y $   & $\sigma_0 \tau_y$  \T\\[3pt]
      \hline
      $C$ & BDI Chiral & $ \sigma_y \tau_y$   & $\hat{d} \cdot \bm{\sigma} \tau_y$  \T\\[3pt]
\end{tabular}
\end{ruledtabular}
\label{tab:summary}
\caption{Summary of symmetry operators and their explicit forms in Examples~1,2.
Particle-hole and time-reversal symmetries are anti-unitary operators which are the product of a unitary operator, acting in particle-hole and spin space, and the complex conjugation operator ${\cal{K}}$.}
\end{table}

{\em Example 2: Ferromagnetic atom (Fe) chain embedded in Pb superconductor:}
Motivated by recent experimental findings \cite{Yazdani_14,Li_14}, as a second illustrative example we consider a simple effective model for a ferromagnetic (Fe) atom chain embedded in a Pb superconductor which can support Majorana modes and topological superconductivity.
The effective mean-field BdG Hamiltonian, which gives a sufficient description of topological superconductivity of the ferromagnetic nanowire assuming that the underlying degrees of freedom in the Pb substrate have been integrated out, can be written as \cite{Dumitrescu,Hui_14,Dumitrescu_14} $H=\sum_{k}\Psi_k^{\dagger}H(k)\Psi_k$ where,
\begin{eqnarray}
\label{eq:FePb}
H(k) & = & (-2t\cos(k)-\mu)\sigma_0\tau_z \\ \nonumber
& + &  \Delta_p\sin(k) \bm{d} \cdot \bm{\sigma}\tau_x + \mathbf{V} \cdot \bm{\sigma} \tau_0.
\end{eqnarray}
In Eq.~(\ref{eq:FePb}) $\Delta_p$ is the magnitude of the induced spin-triplet pairing potential in the ferromagnetic nanowire.
The Zeeman term due to a constant internal magnetization in the ferromagnet, $\mathbf{V}=(V_x,V_y,V_z)$ (which is assumed to be $\bm{V}=(0,0,J)$).

A generic $d$-vector pointing in an arbitrary direction in spin space is written as $\hat{d}=(\sin\theta\cos\phi,\sin\theta\sin\phi,\cos\theta)$ in polar coordinates.
In order to understand the role of mirror symmetry let us first consider Eq.~(\ref{eq:FePb}) with no intrinsic magnetization (i.e $\mathbf{V}=0$).
In this case, the only spin space Pauli matrix  ($\sigma$) appears in the superconducting term and the obvious choice of $\mathcal{M}$ would be $\mathcal{M}=i\hat{d}\cdot\sigma\tau_0$.
Thus the mirror plane is the plane perpendicular to the $\hat{d}$-vector.
Following the procedure discussed previously we define an operator $\mathcal{O}=\Theta\cdot\mathcal{M} = (i\sigma_z\sin\theta\cos\phi+\sigma_0\sin\theta\sin\phi-i\sigma_x\cos\theta)\mathcal{K}=(\hat{d}\cdot\hat{y}+i(\hat{d}\times\hat{y})\cdot\mathbf{\sigma})\mathcal{K}$.
Clearly $\mathcal{O}^2=+1$ and $\mathcal{O}H(k)\mathcal{O}^{-1}=+H(-k)$ and thus fulfills the requirements of a class BDI or pseudo time reversal operator.
The chiral operator $\mathcal{C}$ is $\mathcal{C}=\mathcal{O}\cdot\Xi=\hat{d}\cdot\sigma\tau_y$ for our Hamiltonian.
Indeed, the form of the operators in Example 1 can be understood by the above relation describing the structure of the $\mathcal{O}$ operator.
In the spin-orbit coupled system the direction of the spin-orbit field (remember the Rashba term involved a $\sigma_y$) plays the role of an effective d-vector for the effective p-wave pairing created by the combination of spin-orbit coupling and singlet superconductivity.

Now let us turn on the effective Zeeman field $\mathbf{V}$ and examine what happens to the mirror symmetry.
As long as the only non-zero component of $\mathbf{V}$ is along the $\hat{d}$ vector, mirror symmetry remains intact because the mirror operator still commutes with the Hamiltonian in Eq.~\ref{eq:FePb}.
However this direction of the Zeeman field breaks the chiral symmetry $\mathcal{C}={\cal O}\cdot\Xi$ (as it breaks ${\cal O}={\cal M}\cdot\Theta$) and thus the system belongs to class D where even number of localized Majorana bound states hybridize into finite-energy quasiparticles.
Thus the mirror symmetry alone does \textit{not} protect spatially localized Majorana multiplets at the sample edges.
Let us now introduce a Zeeman field which only lies in the mirror plane which is perpendicular to the $\hat{d}$-vector.
Now the mirror operator no longer commutes with the Hamiltonian $H$ and thus mirror symmetry is broken.
However in this case since $\Theta$ is also broken by the Zeeman field the pseudo TR operator ${\cal O}={\cal M}\cdot\Theta$ remains unbroken. Thus chiral symmetry $\mathcal{C}$ is unbroken and this is a class BDI system as discussed earlier \cite{Dumitrescu,Hui_14,Dumitrescu_14}.

{\em Conclusion:}
In this work we show that topological mirror superconductivity that results from the co-existence of a mirror symmetry and symplectic time-reversal
symmetry may also be viewed as a co-existence of symplectic time-reversal and chiral topological superconductivity in  the BDI symmetry class in one dimension.
The mirror Berry phase \cite{Zhang_13} is found to coincide with the winding number invariant that characterizes the BDI symmetry class in one dimension.
The winding number and other qualitative aspects of the phase such as number of edge Majorana modes, continue to survive even when the symplectic time-reversal
$\Theta$ and mirror-symmetry ${\cal M}$ are broken weakly. Thus the topological mirror phase is adiabatically connected to a BDI phase by infinitesimal
perturbations that break the mirror and time-reversal symmetry while preserving the BDI chiral symmetry ${\cal C}$. Such a perturbation does not have a qualitative
effect such as splitting edge Majorana modes.
We illustrate our point with two examples, namely, a spin-orbit coupled semiconductor nanowire with proximity induced extended $s$-wave pairing potential \cite{Zhang_13}, and the recently discussed experimental system of chains of ferromagnetic atoms on a spin-orbit coupled substrate of Pb superconductor \cite{Yazdani_14,Li_14}.

{\em Acknowledgment:} This work is supported by
AFOSR (FA9550-13-1-0045). J.D.S. would
like to acknowledge the University of Maryland, Condensed
Matter theory center, and the Joint Quantum institute for startup
support.

\end{document}